\documentstyle[12pt]{article}
\begin{document}
\input epsf \renewcommand{\topfraction}{0.8}
\pagestyle{empty}
\vspace*{5mm}
\begin{center}
\Large{\bf Branon search in hadronic colliders} \\
\vspace*{1cm} \large{\bf J.A.R. Cembranos$^{1,\;2}$, }{\bf A.
Dobado$^2$}
{\bf and A. L. Maroto$^2$} \\
\vspace{0.2cm} \normalsize $^1$ Departamento de Estad\'{\i}stica e
Investigaci\'on Operativa III,\\
$^2$ Departamento de  F\'{\i}sica Te\'orica,\\
 Universidad Complutense de
  Madrid, 28040 Madrid, Spain\\
\vspace{0.2cm} \vspace*{0.6cm} {\bf ABSTRACT} \\ \end{center}
\vspace*{5mm} \noindent

In the context of the brane-world scenarios with compactified
extra dimensions, we study the production of brane fluctuations
(branons) in hadron colliders ($p \bar p$, $pp$ and $e^\pm p$) in
terms of the brane tension parameter $f$, the branon mass  $M$ and
the number of branons $N$. From the absence of monojets events at
HERA and Tevatron (run I), we set bounds on these parameters and
we also study how such bounds could be improved at Tevatron (run
II) and the future LHC. The single photon channel is also analyzed
for the two last colliders.


\begin{flushleft} PACS: 11.25Mj, 11.10Lm, 11.15Ex \\

\end{flushleft}


\setcounter{page}{1} \pagestyle{plain}

\textheight 20true cm

\newpage
\section{Introduction}

Since rigid objects do not exist in relativistic theories, it
is clear that brane fluctuations must play a role in the so called
brane world, proposed some years ago by Arkani-Hamed,  Dimopoulos
and  Dvali (ADD scenarios  \cite{ADD}), where the Standard Model
(SM) particles are confined to live in  the world brane and only
gravitons are free to move along the $D>4$ dimensional bulk space
(see \cite{rev} for recent reviews). This fact turns out to be
particularly true when the brane tension scale $f$
($\tau=f^4$ being the brane tension) is much smaller than the
$D$ dimensional or
fundamental gravitational scale $M_D$, i.e., $f<<M_D$. In this
case the only relevant low-energy modes of the ADD scenarios are
the SM particles and branons which are the quantized brane
 oscillations. Branons can be understood as the (pseudo)Goldstone
 bosons corresponding to the spontaneous breaking of translational
 invariance in the bulk space produced by the presence of the brane.
It has been pointed out that branons  could solve some of the problems
of the brane-world scenarios such as the
divergent virtual contributions from the Kaluza-Klein tower at the
tree level or non-unitarity of the graviton production
cross-sections \cite{GB}. As Goldstone bosons, branons are in principle
 massless, but in the cases where the metric of the extra
dimensions cannot be factorized, they can become  massive
\cite{DoMa,BSky}. This is similar to the case of pions which,
being the Goldstone bosons of the spontaneous breaking of chiral
symmetry, acquire some mass due to the explicit breaking of the
symmetry induced by the quark masses.

In previous works the different effective actions have been
obtained, namely: the effective action for the SM fields on the
brane, that for the branon self-interactions and finally that
corresponding to the interaction between SM fields and branons
\cite{DoMa}. In general, this branon effective action can be
parameterized by the number of branons $N$, the tension scale $f$
and the branon masses (for an explicit construction see
\cite{AAGS}). Using the effective action it is possible to obtain
the different Feynman rules, the amplitudes and finally the
cross-sections for branon production from SM particles. In
\cite{ACDM,L3,CrSt} the case of electron-positron colliders has
been considered. By using the Large Electron-Positron Collider
(LEP) data it is possible to set important bounds on the tension
scale and on the branon mass for a given branon number. Other
restrictions have also been set from astrophysical and
cosmological considerations due to the fact that branon dark
matter can present relevant abundances \cite{CDM}.

In this work we study branon production in hadron colliders and
also in electron-proton colliders such as HERA. Most of these
cross sections have been studied by Creminelli and Strumia for the
massless branon case \cite{CrSt}. We reproduce their results and
extend the analysis for an arbitrary branon mass. The paper is
organized as follows: In Sec.II we shortly review the branon
effective action. In Sec.III we consider the case of
proton-(anti)proton colliders like Tevatron or the future Large
Hadron Collider (LHC). In Sec.IV electron(positron)-proton
colliders like HERA are studied. In Sec.V we show the main results
for the relevant examples and in Sec.VI we set the conclusions.

\section{Effective action}

The relevant effective action describing the low-energy
interactions of SM particles and branons was derived in
\cite{ACDM}, where the necessary vertices are detailed. The branon
effective action can be expanded according to the number of branon
fields  appearing in each term:
\begin{equation}
S_{eff}[\pi]=S_{eff}^{(0)}[\pi]+S_{eff}^{(2)}[\pi]+...
\end{equation}
where the zeroth-order term is just a constant and the
second-order is just the free action:
\begin{equation}
S_{eff}^{(2)}[\pi]=\frac{1}{2}\int
d^4x(\delta_{\alpha\beta}\partial_\mu \pi^\alpha\partial^\mu
\pi^\beta-M^2_{\alpha\beta}\pi^\alpha\pi^\beta),
\end{equation}
with $\pi^{\alpha}(x)$  the branon fields where $\alpha=1,2,...,N$
and $M_{\alpha\beta}^2$ is the squared mass matrix which, without
loss of generality, can be assumed to be diagonal. The effective
action for the SM particles and their interactions with branons is
given by
\begin{equation}
S_{SM\pi}=\int d^4x[{\cal L_{SM}}+\frac{1}{8
f^4}(4\delta_{\alpha\beta}\partial_\mu \pi^\alpha\partial_\nu
\pi^\beta- \eta_{\mu\nu}  M^2_{\alpha\beta}\pi^\alpha\pi^\beta
)T_{SM}^{\mu\nu}],
\end{equation}
where ${\cal L_{SM}}$  is the SM Lagrangian and $T_{SM}^{\mu\nu}$
is the SM energy-momentum tensor defined as:
\begin{equation}
T_{SM}^{\mu\nu}=-(g^{\mu\nu}{\cal L_{SM}}+2\frac{\delta{\cal
L_{SM}}}{\delta g_{\mu\nu}})|_{g_{\mu\nu}=\eta_{\mu\nu}},
\end{equation}
where $g_{\mu\nu}$ is some arbitrary metric on the world brane and
$\eta_{\mu\nu}$ is the Minkowski metric.

In this work we are interested in the interactions between quarks
and gluons or photons. Thus, for Dirac fermions with masses $m_q$
belonging to some representation of a gauge group, such as
 $U(1)_{em}$ or $SU(3)_c$, with generators $T^a$, the Lagrangian is
\begin{equation}
{\cal L}_{q}=\bar q ( i \gamma^\mu D_\mu -m_q)q,
\end{equation}
where the covariant derivative is defined as $D_\mu=\partial_\mu-h
A^a_\mu T^a$, $h$ being the appropriate gauge coupling. Thus the
 energy-momentum tensor is given by
\begin{eqnarray}
T_{q}^{\mu\nu}=\frac{i}{4}(\bar q (\gamma^\mu D^\nu+\gamma^\nu
D^\mu)q-(D^\nu \bar q \gamma^\mu+D^\mu \bar q
\gamma^\nu)q)\nonumber
\\
-\eta^{\mu\nu}(i (\bar q \gamma^\rho D^\rho - D_\rho\bar q
\gamma^\rho)q-2 m_q \bar q q ),
\end{eqnarray}
from where it is possible to find vertices such as $\pi\pi\bar q
q$  and $\pi\pi\bar q q A$. For gauge fields $A$ the appropriate
Lagrangian for perturbation theory is:
\begin{equation}
{\cal L_{A}}=-\frac{1}{4}F^{a\mu\nu}F^a_{\mu\nu}+{\cal L_{FP}},
\end{equation}
where as usual $F^a_{\mu\nu}=\partial_\mu A^a_\nu-\partial_\nu
A^a_\mu-hC^{abc}A^b_\mu A^c_\nu$ and ${\cal L_{FP}}$ is the
Fadeev-Popov Lagrangian including the gauge fixing and the ghost
terms. The energy-momentum tensor is:
\begin{equation}
T_{A}^{\mu\nu}=F^a_{\rho\sigma}F^a_{\lambda\theta}(\eta^{\sigma\lambda}\eta^{\rho\mu}
\eta^{\theta\nu}+\frac{1}{4}\eta^{\rho\lambda}\eta^{\sigma\theta}
\eta^{\mu\nu} )+T_{FP}^{\mu\nu},
\end{equation}
from where we can obtain the $\pi\pi AA$, $\pi\pi AAA$ and $\pi\pi
AAAA$ vertices.

Therefore, by using  these energy-momentum tensors and the
effective action above, it is possible to obtain the different
Feynman rules involving branons. One important observation is that
in all the vertices obtained above, branons appear always by
pairs. In fact they interact in a way similar to gravitons since
they couple to the energy momentum tensor. This can be seen by
making the formal identification of the graviton field
$h_{\mu\nu}$ which appear in linearized gravity with
\begin{equation}
h_{\mu\nu}\rightarrow -\frac{1}{\kappa
f^4}(\delta_{\alpha\beta}\partial_\mu\pi^\alpha\partial_\nu\pi^\beta-
\frac{1}{4}\eta_{\mu\nu}M_{\alpha\beta}\pi^\alpha\pi^\beta).
\end{equation}
where $\kappa=4\sqrt{\pi}/M_{P}$ and $M_{P}$ the Planck mass. Of
course the physical meaning is completely different for branons
and  gravitons. In any case branons are expected to be weakly
interacting and then they will scape to detection. Hence their
typical signature will be missing energy and momentum. Since
branons are produced by pairs, the energy spectrum of any
other particle present in the final state will be continuous. 
In the following sections we will study the
production mechanisms relevant for the different hadronic
colliders.

\section{Proton-(anti)proton colliders}

For the case of proton-antiproton colliders like Tevatron, the
most important processes for branon production are quark-antiquark
annihilation or gluon fusion giving a gluon and a branon pair; and
(anti)quark-gluon interaction giving an (anti)quark and a branon
pair. Therefore the expected experimental signal will be in both
cases one monojet $J$ and missing energy and momentum. This is a
very clear signature that in principle can be easily identified.
Another potentially interesting process is the quark-antiquark
annihilation giving a photon and a branon pair. In this case the
signature is one single photon and missing energy and momentum.

\begin{figure}[h]
\hspace*{3cm}
{\epsfxsize=10.0 cm \epsfbox{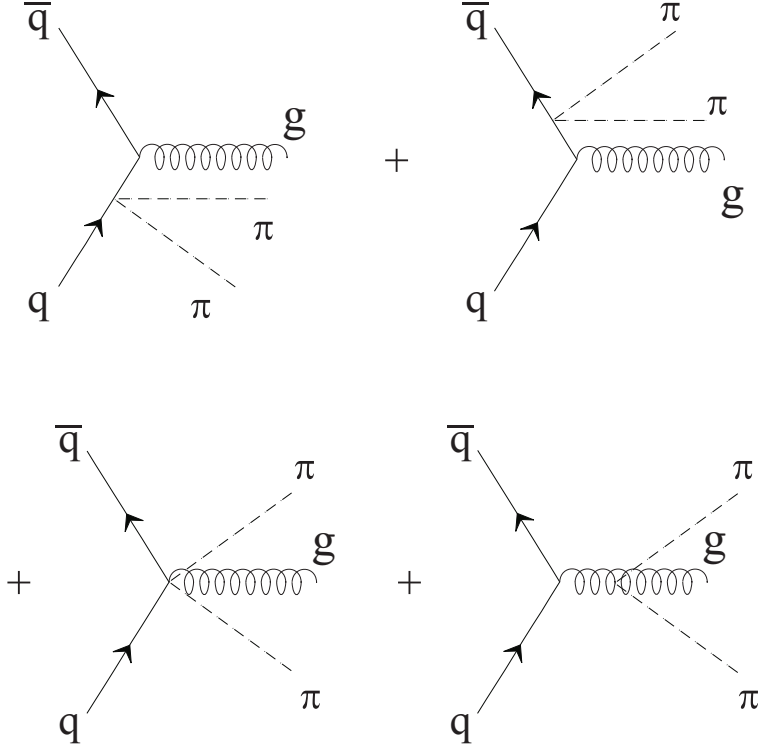}}
\caption{\footnotesize{Feynman diagrams associated to the $q\bar q
\rightarrow g \pi\pi$ subprocess. }}\label{2q}
\end{figure}

\begin{figure}[h]
\hspace*{3cm}
{\epsfxsize=10.0 cm \epsfbox{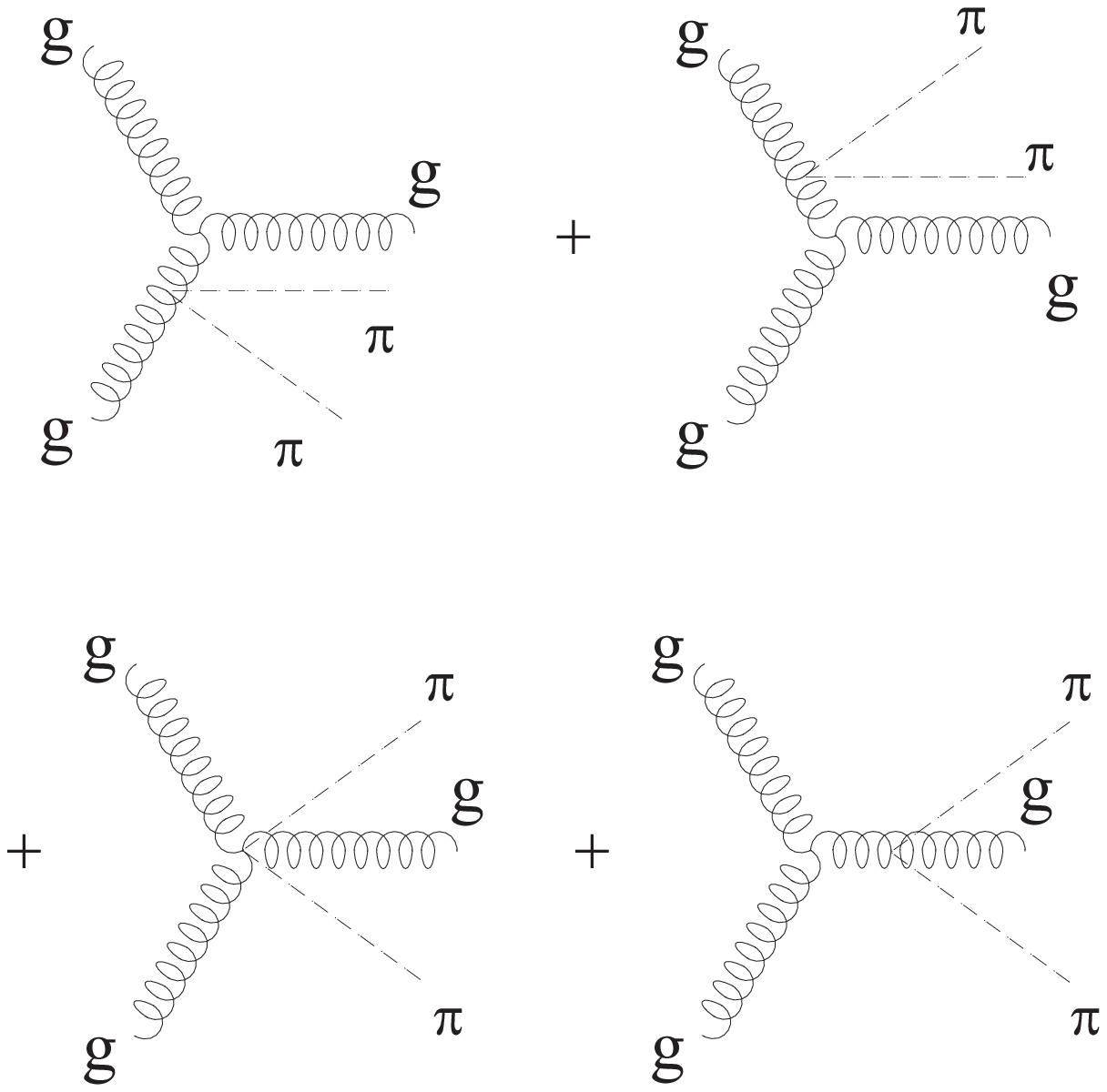}}
\caption{\footnotesize{Feynman diagrams associated to the $g g
\rightarrow g \pi\pi$ subprocess. }}\label{4q}
\vspace*{1cm}
\end{figure}

\begin{figure}[h]
\hspace*{3cm}
{\epsfxsize=10.0 cm \epsfbox{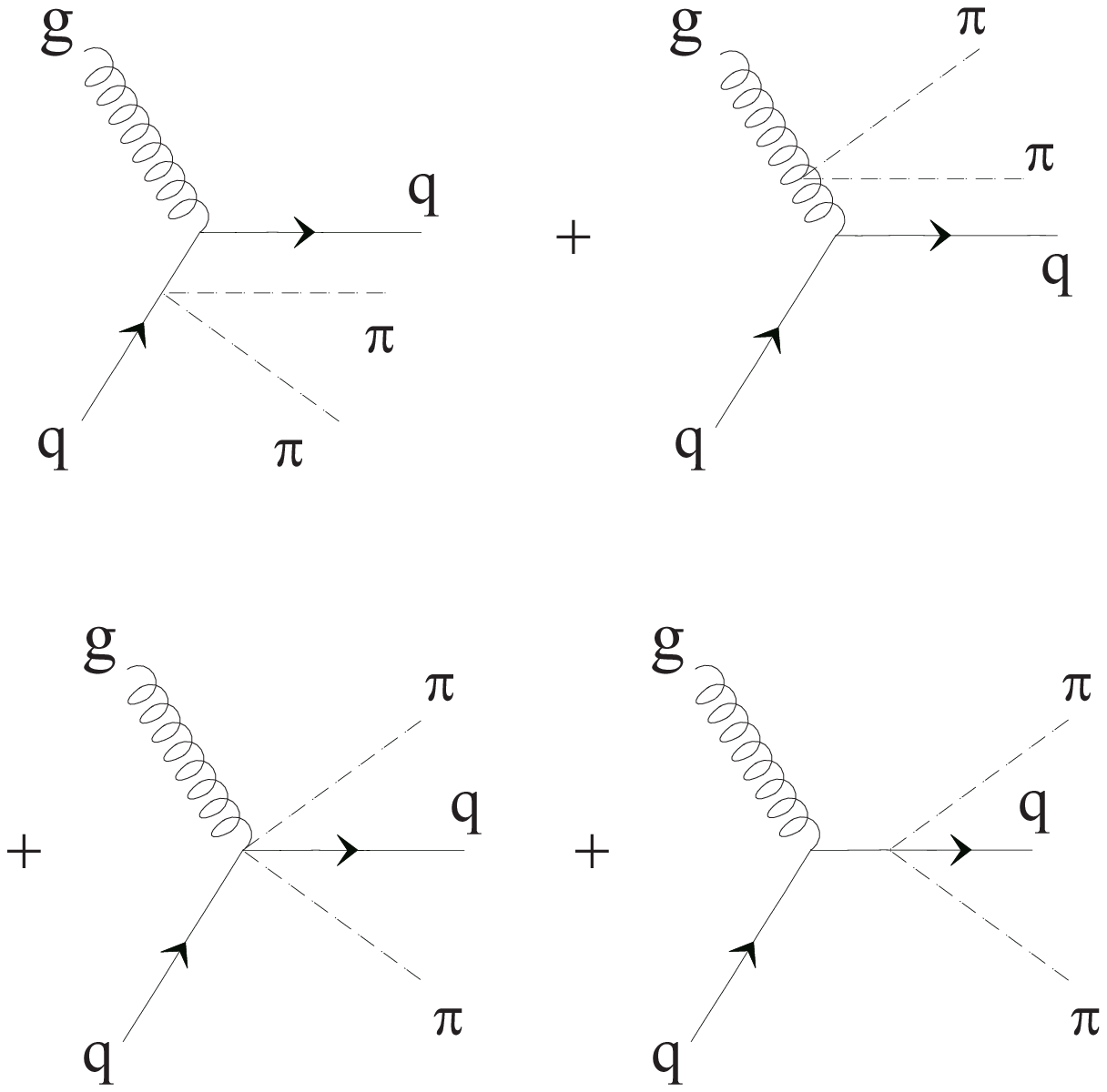}}
\caption{\footnotesize{Feynman diagrams associated to the $ q g
\rightarrow  q \pi\pi $ subprocess. The $\bar q g \rightarrow \bar
q \pi\pi $ subprocess has the same diagrams, but changing the
quark lines by the corresponding antiquark ones.}}\label{3q}
\end{figure}

The Feynman diagrams contributing to the main subprocesses $q\bar
q \rightarrow g \pi\pi$, $g g \rightarrow g \pi\pi$,
$ q g \rightarrow  q \pi\pi $ and  $\bar q g \rightarrow \bar q \pi\pi $
 are shown in Fig. \ref{2q}, Fig. \ref{4q} and Fig. \ref{3q}.

%
From these diagrams and the Feynman rules coming from the
effective action of the previous section, it is possible to obtain
the differential cross section:
\begin{eqnarray}
\frac{d\sigma(q\bar q \rightarrow
g\pi\pi)}{dk^2dt}=\frac{4\alpha_s
N}{3}\frac{(k^2-4M^2)^2}{184320f^8\pi^2\hat
s^3tu}\sqrt{1-\frac{4M^2}{k^2}}
(\hat s k^2 + 4 t u)(2 \hat s k^2+ t^2+u^2),
\end{eqnarray}
where $\hat s\equiv(p_1+p_2)^2$, $t \equiv(p_1-q)^2$,
 $u\equiv(p_2-q)^2$ and $k^2\equiv(k_1+k_2)^2$, $p_1$ and
$p_2$ being the anti-quark and quark four-momenta respectively, $q$ the
gluon four-momentum and $k_1$ and $k_2$ the branon four-momenta.
We have  assumed for the sake of simplicity that all the
branons are degenerated with a common mass $M$ and that
all the quarks are massless. We have also neglected the effects of the
top quark. In addition we
have the well-known relation $\hat s+t+u=k^2$. The contribution to
the total cross section of  the process $p\bar p\rightarrow
g\pi\pi$ coming from this subprocess is given by
\begin{eqnarray}
\label{qqbar}
\sigma_{q\bar q}(p\bar p\rightarrow g\pi\pi)= \int_{x_{min}}^1
dx\int_{y_{min}}^1 dy \sum_q
\bar q_{\bar p}(y;\hat s) q_{ p}(x;\hat s)
\nonumber\\
\int_{k^2_{min}}^{k^2_{max}}  dk^2 \int_{t_{min}}^{t_{max}}
dt\frac{d\sigma(q\bar q \rightarrow g\pi\pi)}{dk^2dt}+...
\end{eqnarray}
where $\bar q_{\bar p}(y;\hat s)$ and $q_{ p}(x;\hat s) $ are the
distribution functions of the anti-quark inside the
antiproton and of the quark of flavor $q$ inside the  proton
respectively, and $x$ and $y$ are the
fractions of  the proton and antiproton energy carried by the
 subprocess quark and antiquark. The different limits of the
 integrals can be written in terms of the cuts used to define
 the total cross-section. For example, in order to be able to
 detect clearly the monojet one must impose a minimal value for its
 transverse energy $E_T$ and a pseudorapidity range given by $\eta_{min}$ and
$\eta_{max}$. Then we have the limits
 $k^2_{min}=4M^2$, $k^2_{max}=\hat s(1-2E_T/\sqrt{\hat s})$ and
 $t_{min(max)}=-(\hat s-k^2)[1+\tanh{(\eta_{min(max)})}]/2$.
 On the other hand $x_{min}=s_{min}/s$ and $y_{min}=x_{min}/x$ where
$s$ is the total center of mass energy squared of the process and
\begin{equation}
s_{min}=2E_T^2+4M^2+2E_T\sqrt{E_T^2+4M^2}.
\end{equation}
In addition the dots in (\ref{qqbar}) represent the contribution
of the case in which the quark comes from the antiproton and the
anti-quark comes from the proton.

The cross-section of the subprocess $g g \rightarrow g \pi\pi$
is given by

\begin{eqnarray}
&&\frac{d\sigma(g g \rightarrow g\pi\pi)}{dk^2dt}=\frac{\alpha_s N
(k^2-4M^2)^2}{40960f^8\pi^2\hat s^3tu}\sqrt{1-\frac{4M^2}{k^2}} \nonumber  \\
&&(\hat s^4+t^4+u^4-k^8+6k^4(\hat s^2+t^2+u^2)-4k^2(\hat
s^3+t^3+u^3)),
\end{eqnarray}
where the Mandelstan variables are defined as in the previous
case, with $p_1$ and $p_2$ being the initial gluon four-momenta,
$q$ the final gluon four-momentum and $k=k_1+k_2$ the total branon
four-momentum. Then the contribution to the total cross section
from the $p\bar p\rightarrow g\pi\pi$ reaction is
\begin{eqnarray}
\sigma_{gg}(p\bar p\rightarrow g\pi\pi)= \int_{x_{min}}^1
dx\int_{y_{min}}^1 dy
g(y;\hat s) g(x;\hat s)   \nonumber\\
\int_{k^2_{min}}^{k^2_{max}}  dk^2 \int_{t_{min}}^{t_{max}}
dt\frac{d\sigma(g g\rightarrow g\pi\pi)}{dk^2dt}.
\end{eqnarray}
Here $g(x;s)$ is the gluon distribution function of the
(anti)proton, $x$ and $y$ are the fractions of the proton and
antiproton energy carried by the initial gluons and the
integration limits remain the same. From the above equations, it
is possible to compute the total cross-section $\sigma( p \bar p
\rightarrow g \pi\pi)$ in terms of the cut in the gluon (monojet)
transverse energy $E_T$.

For the $q g \rightarrow q \pi\pi $ process the cross-section is
given by
\begin{eqnarray}
\frac{d\sigma(q g \rightarrow q\pi\pi)}{dk^2dt}=-\frac{\alpha_s
N}{2}\frac{(k^2-4M^2)^2}{184320f^8\pi^2\hat
s^3tu}\sqrt{1-\frac{4M^2}{k^2}}
(u k^2 + 4 t \hat s)(2 u k^2+ t^2+\hat s^2),
\end{eqnarray}
with  $p_1$ and $p_2$ being the quark and the gluon four-momenta
respectively, $q$ the final state quark  four-momentum and $k_1$
and $k_2$ the branon four-momenta. The Mandelstam variables are
defined as in previous cases. The cross-section for the conjugate
process $\bar q g \rightarrow \bar q \pi\pi$ is exactly the same.
Then the total cross section for the the reaction $p\bar
p\rightarrow q\pi\pi$ is
\begin{eqnarray}
\sigma(p\bar p\rightarrow q\pi\pi)= \int_{x_{min}}^1
dx\int_{y_{min}}^1 dy \sum_q
 g(y;\hat s) q_p(x;\hat s)   \nonumber\\
\int_{k^2_{min}}^{k^2_{max}}  dk^2 \int_{t_{min}}^{t_{max}}
dt\frac{d\sigma(q g\rightarrow q\pi\pi)}{dk^2dt}+...
\end{eqnarray}
%
\begin{figure}[h]
\hspace*{3cm}
{\epsfxsize=10.0 cm \epsfbox{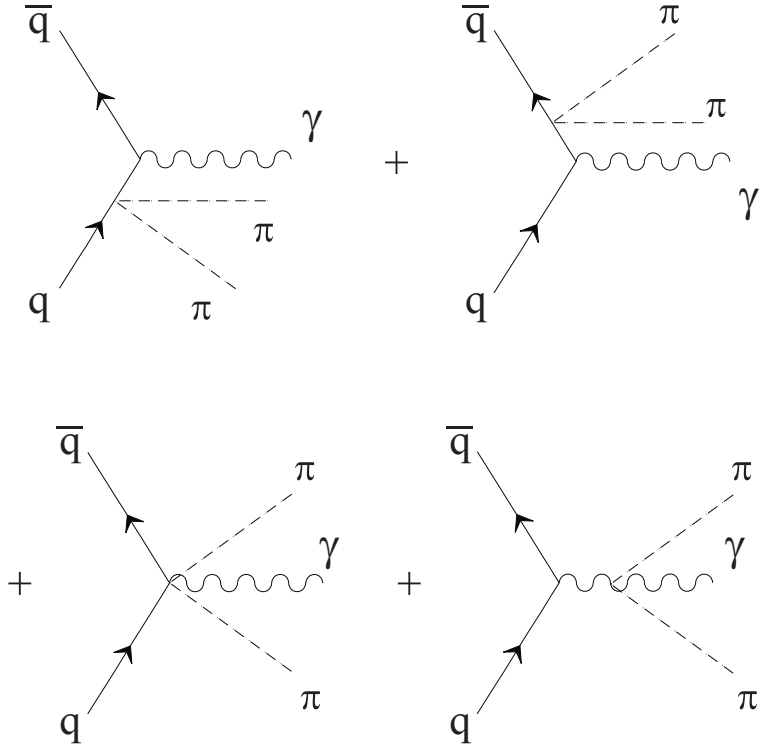}}
\caption{\footnotesize{Feynman diagrams associated to the $q\bar q
\rightarrow \gamma \pi\pi$ subprocess. }}\label{1q}
\end{figure}
%
In this equation $x$ and $y$ are the fractions of  the proton and
antiproton energy carried by the
 subprocess quark and gluon. The different integration limits are
 defined as in the previous case in terms of the minimal
 transverse energy of the quark (monojet) $E_T$ and the dots
 refer to the case where the initial gluon is coming from the
 proton and the quark is coming from the antiproton.
In addition we have the contribution from the conjugate case where we take
an anti-quark from the proton and a gluon from the antiproton and
conversely. This amount just to a factor of two.

From all the above equations it is possible to compute the total
cross-section $\sigma( p \bar p \rightarrow J \pi\pi)$ in terms of
the cut in the jet transverse energy $E_T$.

For the subprocess $q\bar q \rightarrow \gamma \pi\pi$ the
cross-section is given by
\begin{eqnarray}
\frac{d\sigma(q\bar q \rightarrow
\gamma\pi\pi)}{dk^2dt}=\frac{Q_q^2\alpha
N(k^2-4M^2)^2}{184320f^8\pi^2\hat s^3tu}\sqrt{1-\frac{4M^2}{k^2}}
(\hat s k^2 + 4 t u)(2 \hat s k^2+ t^2+u^2).
\end{eqnarray}
Here the notation is similar to the $q\bar q \rightarrow g \pi\pi$
case with the obvious differences in couplings, color and charge
factors. Thus
\begin{eqnarray}
\sigma(p\bar p\rightarrow \gamma\pi\pi)= \int_{x_{min}}^1
dx\int_{y_{min}}^1 dy \sum_q
\bar q_{\bar p}(y;\hat s) q_{ p}(x;\hat s)   \nonumber\\
\int_{k^2_{min}}^{k^2_{max}}  dk^2 \int_{t_{min}}^{t_{max}}
dt\frac{d\sigma(q\bar q \rightarrow \gamma\pi\pi)}{dk^2dt}+...
\end{eqnarray}
All the previous discussion about branon production in $p \bar p$
reactions can be easily translated to the $p p$ case. The only
point is to change the antiproton distribution functions of the
different partons by the corresponding proton ones.

\section{Electron(positron)-proton colliders}

\begin{figure}[h]
\hspace*{3cm}
{\epsfxsize=10.0 cm \epsfbox{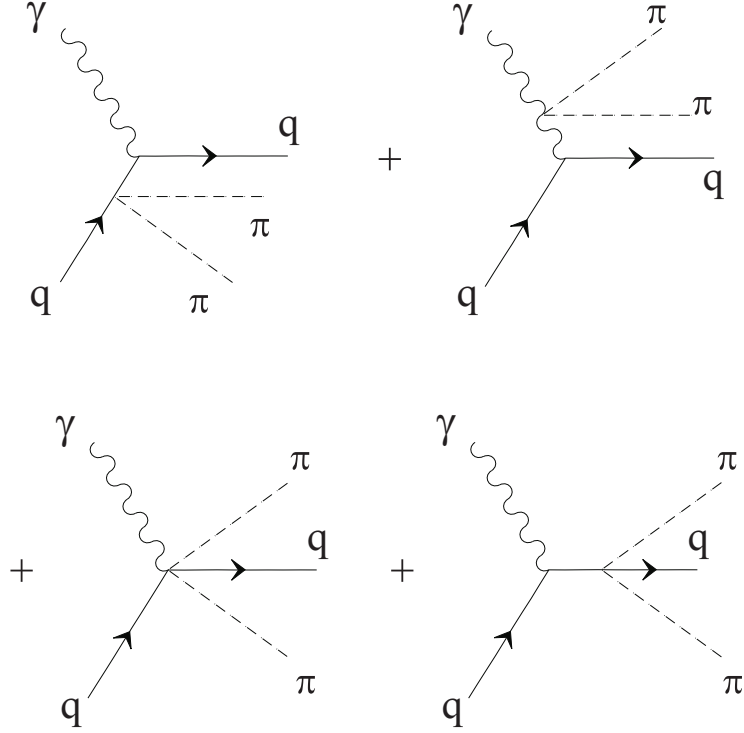}}
\caption{\footnotesize{Feynman diagrams associated to the $q\gamma
\rightarrow q  \pi\pi$ subprocess. The $\bar q\gamma
\rightarrow \bar q  \pi\pi$} subprocess has the same
diagrams but changing the quark lines by the corresponding
antiquark ones.}\label{5q}
\end{figure}

For electron(positron)-proton colliders like HERA, the most
interesting branon creating process  is branon photoproduction,
where a photon emitted by the electron(positron) interacts with a
quark(antiquark) from the proton giving a quark (antiquark) and a
branon pair. Thus the experimental signature is again one monojet
$J$ plus missing energy and momentum. The relevant Feynman
diagrams are shown in Fig. \ref{5q} and the corresponding
differential cross-section for the subprocess $\gamma q
\rightarrow q\pi\pi$ is
\begin{eqnarray}
\frac{d\sigma(\gamma q\rightarrow
q\pi\pi)}{dk^2dt}=-\frac{3Q_q^2\alpha N
(k^2-4M^2)^2}{184320f^8\pi^2{\hat s}^3tu}\sqrt{1-\frac{4M^2}{k^2}}
(uk^2+4 t {\hat s}^2)(2uk^2+t^2+{\hat s}^2),
\end{eqnarray}

\noindent where ${\hat s}=(p+q)^2, t=(p-k)^2$ and $u=(q-k)^2$, $p$
being the photon, $q$ the proton quark, $q'$ the final quark and
$k$ the total branon momenta respectively. The total cross-section
for the process $e^{\pm}p \rightarrow q \pi\pi$ is given by
\begin{eqnarray}
\sigma(e^{\pm}p \rightarrow q \pi\pi)= \int_{x_{min}}^1
dx\int_{y_{min}}^1 dy \sum_q
 F(y) q_{ p}(x;\hat s)   \nonumber\\
\int_{k^2_{min}}^{k^2_{max}}  dk^2 \int_{t_{min}}^{t_{max}}
dt\frac{d\sigma(\gamma q\rightarrow q\pi\pi)}{dk^2dt},
\end{eqnarray}
$x$ and $y$ are defined in this case as $q=xP_p$ and $P=yP_e$ with
$P_p$ and $P_e$ being the proton and electron(positron) momenta
respectively. Thus at high energies compared with the proton mass
${\hat s}=xys$ where $s=(P_e+P_p)^2$. The integral limits
$y_{min}$, $x_{min}$, $k^2_{min,max}$ and $t_{min,max}$ are
defined like in the proton-(anti)proton collider case.

The photon spectrum $F(y)$ can be obtained from the well-known
Weizs$\ddot{a}$cker-Williams approximation \cite{WW}:
\begin{equation}
F(y)=\frac{\alpha}{2\pi y}[1+(1-y)^2]\log\frac{s'}{4m_e^2},
\end{equation}
with $s'=xs$ and $m_e$ being the electron mass.

The cross-section $\sigma(e^{\pm}p\rightarrow \bar q\pi\pi)$ can
be obtained in a similar way. Then the total contribution to
monojet plus missing energy and momentum production for large
enough $E_T$ coming from branons can be written as the sum of
$\sigma(e^{\pm}p \rightarrow q \pi\pi)$ and $\sigma(e^{\pm}p
\rightarrow \bar q \pi\pi)$.

\section{Results}

By using the cross-sections shown in the previous sections it is
possible to compute the expected number of branon pairs produced
in the different hadron colliders in terms of the brane tension
parameter $f$, the branon mass $M$ and the number of branons $N$.
To this end we have used the distribution functions which can be
found in \cite{partons}. The values
of the electromagnetic and
strong couplings have been taken at the electroweak boson
masses, namely $\alpha=0.0781$ and $\alpha_s=0.1171$.
However our final results do not depend too much on the precise
value of these couplings. In fact our main source of error is the
use of an effective action to describe the SM particles and branon
interactions since, in principle, this is only guaranteed for
energies well below $4\pi f$.
\begin{figure}[h]
{\epsfxsize=12.0 cm \epsfbox{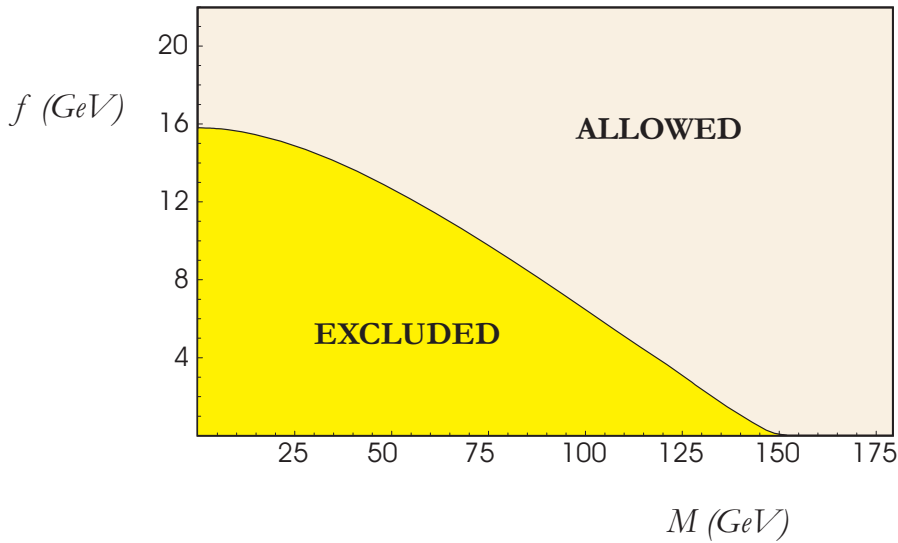}}
\caption{\footnotesize{Exclusion region at the 95 $\%$ c.l.
in the parameter space
$f-M$ for $N=1$ from ZEUS data on jet production. }}\label{hera1}
\end{figure}

As discussed in the introduction, our main goal in this work is to
study the bounds that can be set on the $f$, $M$ and $N$
parameters coming from hadron colliders. We will present all our
limits at the 95\% confidence level. In particular, for the
electron(positron)-proton case, HERA is the most relevant
experiment. In fact, the ZEUS collaboration has studied the jet
production in charged current deep inelastic $e^+p$ scattering.
Its results are perfectly compatible with the SM background and
therefore, we can set some bounds on the branon production and
hence on the $f$, $M$ and $N$ parameters. These data were taken
from 1995 to 2000 at a maximum CM energy of $318$ GeV. The total
integrated luminosity was $110.5$ pb$^{-1}$ and the cuts on the
pseudorapidity and the transverse energy were $-1\leq\eta\leq 2$
and $E_T \geq 14$ GeV (see \cite{ZEUS} for more details). By using
the same cuts with our cross-sections for monojet plus a branon
pair production, we find the bound $f>16N^{1/8}$ GeV for massless
branons. For a branon mass larger than $152$ GeV there is no
restrictions on the $f$ value because of kinematical reasons. For
the intermediate $M$ values the bounds obtained can be seen in
Fig. \ref{hera1} where we have assumed $N=1$. For other $N$ values
one just has to take into account that the bound scales like
$N^{-1/8}$ since all the cross-sections are proportional to
$Nf^{-8}$.
\begin{figure}[h]
{\epsfxsize=12.0 cm \epsfbox{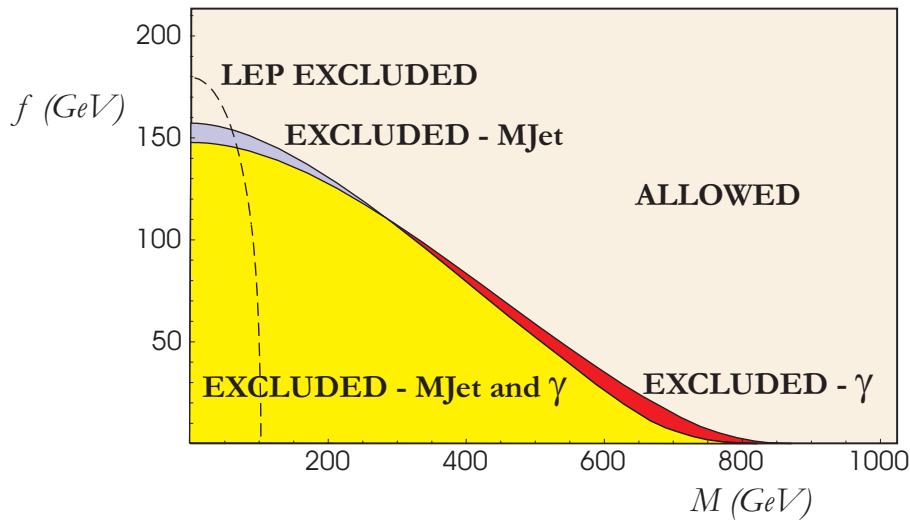}}
\caption{\footnotesize{Exclusion region $f-M$ at the 95 $\%$ c.l.
for $N=1$ with $D\emptyset$ data in the monojet channel, and with
CDF data in the single photon channel. The dashed line
corresponds to the LEP-II limits obtained by the L3 collaboration
using single-photon data \cite{L3}}}\label{tevatron1}
\end{figure}

In the $p\bar p$ case the most relevant experimental information
so far is the one obtained at the Tevatron (Run I). The
$D\emptyset$ detector has studied the monojet channel and CDF the
single photon one. As far as the number of events found in both
cases is compatible with the SM background, we can set new bounds
on the branon theory parameters. For light branons the most
important bound comes from the $D\emptyset$ data. These data were
taken from 1994 to 1996 at a CM energy of $1.8$ TeV and correspond
to an integrated luminosity $78.8\pm 3.9$ pb$^{-1}$. The cuts on
the pseudorapidity and the transverse energy were $|\eta|\leq 1$
and $E_T \geq 150$ GeV (see \cite{D0} for the details of the
analysis). The total number  of monojets observed was $38$ and the
expected number from the SM plus cosmic rays events was $38\pm
9.6$. By using our cross sections for monojet plus a branon pair
production with these cuts we get the bound $f>157N^{1/8}$ GeV for
light branons. The restrictions for $f$ improve up to a branon
mass of $822$ GeV. For the intermediate $M$ values the bounds
obtained can be seen in Fig. \ref{tevatron1} for $N=1$.
\begin{figure}[h]
{\epsfxsize=12.0 cm \epsfbox{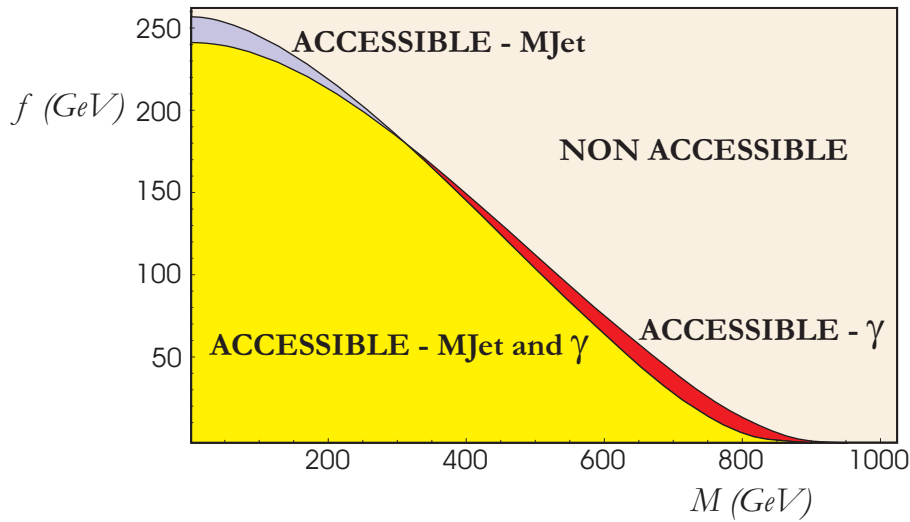}}
\caption{\footnotesize{Sensitivity estimation at the 95 $\%$ c.l.
 for the second run
of Tevatron in the parameter space $f-M$ for
$N=1$.}}\label{tevatron2}
\end{figure}

In a similar way we can use the CDF data on single photon
production. In this case the total luminosity collected was $87\pm
4$ pb$^{-1}$ and the pseudorapidity cut was $|\eta|\leq 1$. For
the transverse photon energy several cuts were considered (for
example 55 GeV at the $75\%$ efficiency). The total expected
background for this process was $11.0\pm 2.2$, without taking into
account the QCD contribution (see \cite{CDF} for the details of
the analysis), and the number of events found was $11$. Comparing
this result with our computations for photon plus one branon pair
production, we find the bound $f>148N^{1/8}$ GeV for massless
branons and no bound for $M$ larger than $872$ GeV. The bound
obtained for the rest of the cases is shown also in Fig.
\ref{tevatron1}.
\begin{figure}[h]
{\epsfxsize=12.0 cm \epsfbox{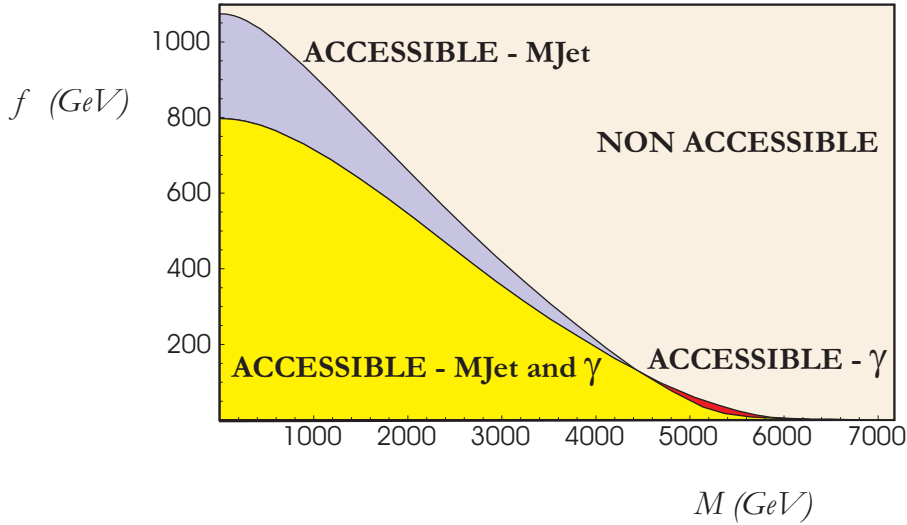}}
\caption{\footnotesize{Sensitivity estimation at the 95 $\%$ c.l.
for the LHC in the
$f-M$ plane for $N=1$.}}\label{lhc1}
\end{figure}

In addition to this analysis corresponding to the Tevatron data
(Run I), it is also interesting to make some estimation about the
bounds that could be set from future experiments such as Tevatron
(Run II) and the LHC. In the case of the Tevatron (Run II), which
is already in progress, the main novelties are a CM energy which
equals $1.96$ TeV and an expected integrated luminosity ${\cal
L_{II}}$ at the end of the run of about $200$ pb$^{-1}$. The
detectors are also improved so that the pseudorapidity cuts can be
taken as $|\eta|\leq 3$ for $D\emptyset$ and $|\eta|\leq 3.6$ for
CDF. This would result in a factor of $\sqrt{{\cal L_{II}}/{\cal
L_{I}}}$ on the statistical significance when compared to the Run
I, with integrated luminosity ${\cal L_{I}}$, provided that the CM
energy and the cuts were the same. For massless branons, the bound
on $f$ scales as the CM energy $E_{CM}^{3/4}$. Even more important
is the possibility of exploring higher branon masses, since the
kinematical limit is given by $M_0=\sqrt{E_{CM}^2-2E_{CM}E_T}/2$.
In Fig. \ref{tevatron2} we show the expected bounds from the Run
II in the $f-M$ plane, again for $N=1$.

The LHC will produce $p p $ collisions at a CM energy of $14$ TeV
and the integrated luminosity will be something about $10^5$
 pb$^{-1}$. In order to estimate the bounds on the $f$, $M$ and $N$ parameters
 that will be possible to obtain at the LHC, we have proceeded in a
 similar way as in the Tevatron case, with the obvious changes in
 the distribution functions due to the fact that now we are
 dealing with $p p $  instead of $p \bar p$ collisions. We have
 kept the same cuts except for the transverse energy which has been
 corrected in order to maintain  the same proportion relative to
 the CM energy. Again the best  bounds for $f$ come from monojet
 production, which for $M=0$ turns out to be $f>1075 N^{1/8}$ GeV. For
 low $f$ the best bound for $M$ is given by the single photon
 channel ($M_0=6781$). The LHC sensitivity for other values in the
 $f-M$ plane can be found in Fig. \ref{lhc1} for $N=1$.

\section{Conclusions}

In this work we have studied the flexible brane-world scenario,
where the brane tension scale $f$ is much smaller than the
fundamental $D$-dimensional gravitational scale $M_D$. In this
case, the relevant low-energy degrees of freedom are the SM
particles and the brane fluctuations or branons. From the
corresponding effective action, we have calculated the relevant
cross-sections for different branon searches in hadronic
colliders. We have used the information coming from HERA and the
first Tevatron Run in order to get different exclusion plots on
the branon mass $M$ and the tension scale $f$ plane for a given
branon number $N$. Monojet production turns out to be the most
efficient process for light branons, whereas the single photon
channel is the most important one for heavy branons.
\begin{table}[h]
\centering \small{
\begin{tabular}{|c|ccccccc|}
\hline\hline Exper.
&
$\sqrt{s}$(TeV)& ${\mathcal
L}$(pb$^{-1}$)&$E_{T}$(GeV)&$\eta_{min,max}$
&$\sigma_0$(GeV$^{-2}$)&$f_0$(GeV)&$M_0$(GeV)\\
\hline
%
%
HERA$^{\,1}
$& 0.318 & 110.5 & 14 & -1, 2 & $7.0\, 10^{-7}$ & 16 & 152
\\
Teva-I$^{\,1}
$& 1.8 & 78 & 150 & -1, 1 & $6.3\, 10^{-10}$ & 157 & 822
\\
Teva-I$^{\,2}
$ & 1.8 & 87 & 55 & -1.1, 1.1 & $1.3\,10^{-10}$ & 148 & 872
\\
Teva-II$^{\,1}
$& 1.96 & $10^3$ & 150 & -3, 3 & $3.2\, 10^{-10}$ & 256 & 902
\\
Teva-II$^{\,2}
$& 1.96 & $10^3$ & 55 & -3.6, 3.6 & $7.0\, 10^{-11}$ & 240 & 952
\\
LHC$^{\,1}
$& 14 & $10^5$ & 1000 & -3, 3 & $1.8\, 10^{-11}$ & 1075 & 6481
\\
LHC$^{\,2}
$& 14 & $10^5$ & 430 & -3.6, 3.6 & $3.8\,10^{-12}$ & 797 & 6781
\\
\hline\hline
\end{tabular}
} \caption{\footnotesize{Summary of the main characteristics of
the analysis performed for hadronic colliders. All the results are
presented at the $95\;\%$ c.l. We have studied two channels: the
one marked with an upper index $^1\,$ is related to monojet
production, whereas the single photon is labelled with an upper
index $^2\,$. We considered four different experiments: HERA, the
I and II Tevatron runs and the LHC. Obviously the data
corresponding  to the two last experiments are estimations,
whereas the first two analysis have been performed with real data.
$\sqrt{s}$ is the center of mass energy associated to the total
process; ${\mathcal L}$ is the total integrated luminosity;
$E_{T}$ is the transverse energy cut; $\eta_{min,max}$, the
pseudorapidity limits; $\sigma_0$ is the estimation for the cross
section sensitivity limit; $f_0$, the bound in the brane tension
scale for one massless branon ($N=1$) and $M_0$ the limit on the
branon mass for $f=0$.}}
\label{tabHad}
\end{table}

We have also extended the analysis to future hadronic colliders.
The corresponding sensitivity regions for the second Tevatron run
and the LHC have also been plotted (see Table \ref{tabHad} for a
summary of the analysis).

These analysis improve those already done for electron-positron
colliders for heavy branons, whereas for light
branons, the results are similar \cite{ACDM,L3,CrSt}. 
The Tevatron (run I) limit
$M_0=872$ GeV can be compared to the analogous limit from LEP II
$M_0= 103$ GeV \cite{L3}. According to the previous estimations,  
the Tevatron run II could
also improve the bound  $f_0=180$ GeV obtained by LEP-II. 
On the other hand, LHC could detect branons
up to a mass of several TeV ($M_0=6781$ GeV) improving even the
CLIC prospects ($M_0\simeq 2500$ GeV) \cite{ACDM}.

The study of branons in colliders can be complemented with other
bounds  coming from astrophysics and cosmology (see Fig. 10). 
In fact, as shown
in \cite{CDM}, the branon relic abundance can have cosmological
consequences. Other issues related to branon phenomenology, such as
their radiative corrections to the SM processes, or their distinctive
signatures at colliders with respect to the KK gravitons will be
analyzed elsewhere.

\begin{figure}[h]
\centerline{\epsfxsize=12cm\epsfbox{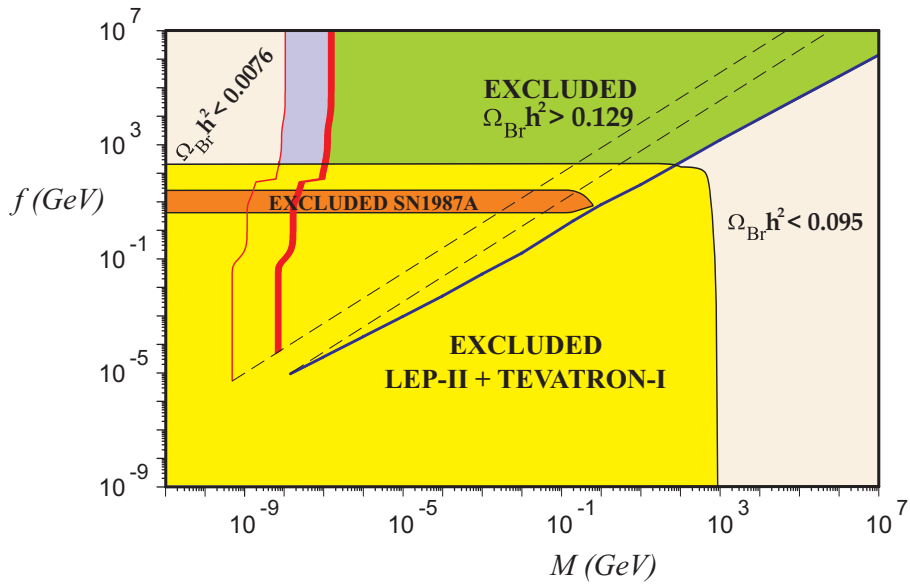}}

\noindent
\caption{{\footnotesize  Relic abundance in the $f-M$ plane for a 
model with one branon of
mass: $M$. The two lines on the left correspond to the 
$\Omega_{Br}h^2=0.0076$ and 
$\Omega_{Br}h^2=0.129 - 0.095$ curves for hot-warm relics, 
whereas the right line
corresponds to the latter limits for cold relics (see \cite{CDM} 
for details). 
The lower area is excluded by single-photon processes at LEP-II 
\cite{L3} together 
with monojet signal at Tevatron-I. The astrophysical 
constraints are less
restrictive and they mainly come from supernova cooling by branon 
emission 
\cite{CDM}.}}
\end{figure}
\vspace*{0.2cm}

 \vspace{.5cm}
{\bf Note added:} After this paper was completed, we were informed
by M. Spiropulu that CDF collaboration had performed a monojet
study \cite{CDF2} which could improve the bounds in a more 
detailed analysis.

 {\bf Acknowledgements:} We would like to thank J. Terr\'on, M.
V\'azquez and M. Spiropulu for useful comments 
and experimental information. This
work has been partially supported by the DGICYT (Spain) under the
project numbers FPA2000-0956 and BFM2002-01003.   \\


\end{document}